\documentclass[conference]{IEEEtran}
\IEEEoverridecommandlockouts

\usepackage{cite}
\usepackage{amsmath,amssymb,amsfonts}
\usepackage{array}
\usepackage{algorithmic}
\usepackage{graphicx}
\usepackage{textcomp}
\usepackage{xcolor}
\usepackage{float}
\usepackage{booktabs}
\usepackage{multirow}
\usepackage{epstopdf}
\usepackage{subfigure}
\usepackage{algorithm}
\usepackage{makecell}

\def\BibTeX{{\rm B\kern-.05em{\sc i\kern-.025em b}\kern-.08em
    T\kern-.1667em\lower.7ex\hbox{E}\kern-.125emX}}
    
\DeclareRobustCommand*{\IEEEauthorrefmark}[1]{%
    \raisebox{0pt}[0pt][0pt]{\textsuperscript{\footnotesize\ensuremath{#1}}}}

\begin{document}

\title{MiniFed : Integrating LLM-based Agentic-Workflow for Simulating FOMC Meeting}

\author{
    \IEEEauthorblockN{Sungil Seok\IEEEauthorrefmark{1}, Shuide Wen\IEEEauthorrefmark{1}, Qiyuan Yang\IEEEauthorrefmark{2}, Juan Feng\IEEEauthorrefmark{3*}, Wenming Yang\IEEEauthorrefmark{1*}}
    \IEEEauthorblockA{\IEEEauthorrefmark{1} Shenzhen International Graduate School, Tsinghua University, Shenzhen, China}
    \IEEEauthorblockA{\IEEEauthorrefmark{2} SKEMA Business School, Nanjing Audit University, Nanjing, China}
    \IEEEauthorblockA{\IEEEauthorrefmark{3} School of Economics and Management, Tsinghua University, Shenzhen, China}

    \IEEEauthorblockA{shi-cy22@mails.tsinghua.edu.cn, yangqiyuan@stu.nau.edu.cn,\\ \{wenshuide, yang.wenming\}@sz.tsinghua.edu.cn, fengjuan@sem.tsinghua.edu.cn}
}
\maketitle

\begin{abstract}
The Federal Funds rate in the United States plays a significant role in both domestic and international financial markets. However, research has predominantly focused on the effects of adjustments to the Federal Funds rate rather than on the decision-making process itself. Recent advancements in large language models (LLMs) offer a potential method for reconstructing the original FOMC meetings, which are responsible for setting the Federal Funds rate. In this paper, we propose a five-stage FOMC meeting simulation framework, \textit{MiniFed}, which employs LLM agents to simulate real-world FOMC meeting members and optimize the FOMC structure. This framework effectively revitalizes the FOMC meeting process and facilitates projections of the Federal Funds rate. Experimental results demonstrate that our proposed MiniFed framework achieves both high accuracy in Federal Funds rate projections and behavioral alignment with the agents' real-world counterparts. Given that few studies have focused on employing LLM agents to simulate large-scale real-world conferences, our work can serve as a benchmark for future developments.

\end{abstract}


\section{Introduction}
Despite a slight decline in its share until 2021, the US dollar continues to reign as the most crucial currency in foreign exchange reserves (FXR), reinforcing its pivotal role in global finance\cite{arslanalp2022stealth}. This prominence has generated substantial interest in the adjustments of US interest rates. While extensive research exists on the effects of these adjustments on foreign economies\cite{iacoviello2019foreign}, international capital flows\cite{warnock2009international}, stock markets\cite{huang2016us}, and even the cryptocurrency market\cite{karau2023monetary}, the process by which US monetary policies are formulated remains less examined. 

The intricacies of the policy-making process play a significant role in the scarcity of research in this domain. The Federal Open Market Committee (FOMC) is tasked with setting the Federal Funds rate, which directly impacts US interest rates. While the FOMC discloses all pertinent materials and the complete discussion process for determining the target interest rate range five years post-meeting, effectively and accurately reconstructing this entire process continues to pose a significant challenge.

However, the breakthroughs in AI technologies, especially in the area of generative models, make addressing this challenge feasible. Generative models, particularly Large Language Models (LLMs) such as ChatGPT\footnote{https://openai.com/index/chatgpt/}, have proven to be exceptionally effective as agents capable of producing realistic simulations of human behavior. These models adapt dynamically to their changing experiences and environment under specific agent architectures\cite{park2023generative}. A growing body of research explores the potential of LLM-based multi-agent systems for real-world scenario applications, including simulations of the US Supreme Court\cite{hamilton2023blind}, classroom environments\cite{zhang2024simulating}, everyday life\cite{park2023generative}, and market research participant role-playing\cite{li2024frontiers}. This extensive research underscores the capabilities of LLM-based multi-agent systems and has motivated us to develop and utilize our own architecture to accurately reconstruct the decision-making process for US interest rate adjustments.

In this paper, we introduce \textit{MiniFed}, our innovatively designed LLM-based multi-agent system that leverages the capabilities of the newly developed ChatGPT-4o mini model in conjunction with our custom FOMC simulation architecture. This system is designed to accurately replicate each FOMC meeting from 2018. Experimental results indicate that our architecture excels in both predictive accuracy, alignment and content comprehension and generation.

The paper is structured as follows: we will begin with a review of related works, followed by a detailed introduction of our MiniFed architecture, the experimental setup, the results of our experiments and further reflections and discussions.

In summary, the paper makes the following contributions:
\begin{itemize}
    \item{We reconstructed the FOMC meeting participants by modeling their socio-demographic and personality attributes to align with their real-world counterparts, thereby demonstrating the accuracy of our agent pre-definition method.}
    
    \item{We proposed our five-stage \textit{MiniFed} architecture, which captures the most important aspects of the FOMC meeting, effectively facilitating communication among the participant agents and eliciting the final monetary policies.}
    
    \item{Given that few prior studies have focused on reconstructing large-scale real-world conferences, our work can serve as a benchmark for subsequent research.}
    
\end{itemize}

\begin{figure*}[htbp]
\centering
\includegraphics[scale=0.3]{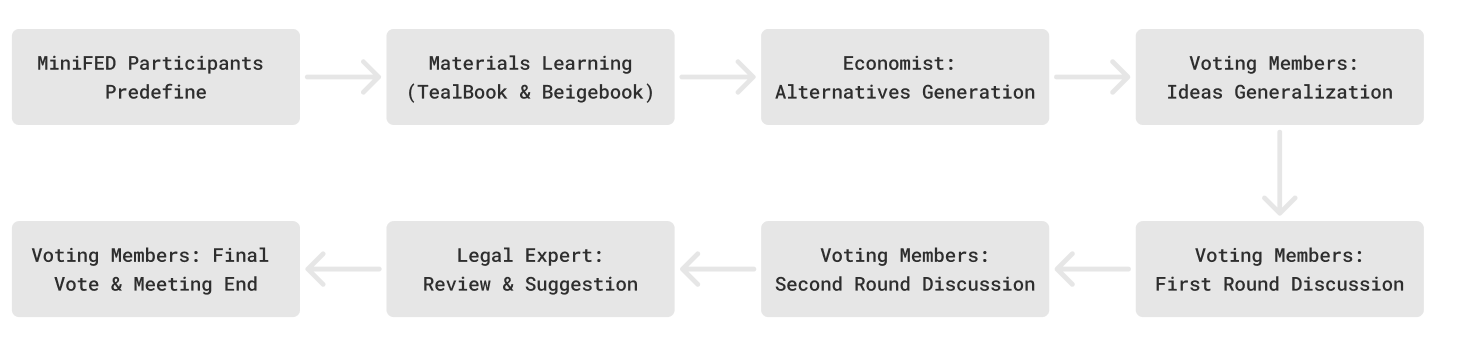}
\caption{MiniFed Pipeline}
\label{figure1}
\end{figure*}

\section{Related Works}
In this section, we review prior literature on the reflection and application of FOMC materials, credible proxies for human behavior, and LLM-based multi-agent frameworks for real-world applications. We demonstrate that current LLM models are capable of simulating complex interactive human behaviors, such as those observed in FOMC meetings.

\subsection{FOMC Related Researches}\label{AA}
The Federal Open Market Committee (FOMC) plays a crucial role in shaping the United States' monetary policy to promote economic stability and growth. Its primary duties include setting interest rates, conducting open market operations to regulate the money supply, and assessing economic conditions to achieve goals like controlling inflation and maximizing employment. Its importance lies in its ability to influence national and global financial markets, impact borrowing costs, and steer the economy toward sustainable growth through informed policy decisions. 

The FOMC disseminates a comprehensive array of materials throughout its meeting process, including transcripts (detailed records of meeting proceedings), tealbooks (economic analyses and descriptions of policy alternatives), agendas (lists of topics to be addressed at each meeting), minutes (summaries of issues discussed), and beige books (information on current economic conditions by district). Although the materials provided by the FOMC are detailed and exhaustive, research on the FOMC remains in its early stages.

Existing literature primarily focuses on empirical analyses of the effects of FOMC monetary policy announcements on global asset prices\cite{hausman2011global} and U.S. financial market responses\cite{cieslak2019stock}\cite{rosa2013financial}. Additionally, some researchers argue that informed traders in the financial market are able to trade effectively before the FOMC releases its monetary policy announcements. Consequently, studying financial market turbulence prior to the publication of the FOMC's target rate is also considered valuable\cite{du2018informational}\cite{lucca2015pre}.

Recently, as deep learning technologies have begun to emerge, many researchers have sought to employ deep learning–enhanced methods to analyze contextual sentiments in documents released by the FOMC. Tsang et al.\cite{tsang2023agree} developed a deep learning model based on a self-attention mechanism to measure disagreement among members in each FOMC meeting, utilizing records of dissents on FOMC votes and meeting transcripts from 1976 to 2017. Additionally, with the growing popularity of fine-tuning methods for large language models (LLMs), Gössi et al.\cite{gossi2023finbert} fine-tuned the pre-trained FinBERT model to analyze the sentiments present in FOMC Minutes materials. However, overall research on deep learning–based approaches to FOMC materials remains in its infancy.

\subsection{LLM-Based Multi-Agent Frameworks }
As the field of artificial intelligence rapidly expands, researchers are exploring multi-agent systems in which multiple AI entities collaborate to achieve common objectives and address complex real-world challenges\cite{dohan2022language}\cite{talebirad2023multi}. One notable example is CAMEL\cite{li2023camel}, which introduces a communicative multi-agent framework designed to reduce the need for human user input prompts and provides a scalable approach for examining the cooperative behaviors and capabilities of multi-agent systems.

Another intriguing area in this field involves utilizing large language model (LLM) agents to simulate real-world scenarios. LLMs, particularly GPT-4\cite{bubeck2023sparks}, have demonstrated near-human-level performance across a variety of challenging tasks, including mathematics\cite{zhang2024mathverse}, coding\cite{nam2024using}, vision\cite{wang2024visionllm}, medicine\cite{thirunavukarasu2023large}, law\cite{cui2023chatlaw}, psychology\cite{yang2024llm}, and more. Consequently, developing methods to enable LLMs to behave like real humans within simulated environments and observing their interactions and performance represents a compelling research direction.

One notable study involves simulating human community life and monitoring interactions, behaviors, and dialogues in a game environment by leveraging generative agents and a novel agent architecture\cite{park2023generative}. Additionally, other studies have focused on simulating various real-world scenarios by deploying LLM agents, such as school environments\cite{zhang2024simulating} and travel trajectories\cite{li2024more}.

Specifically, simulating human roles within specific scenarios is important because it provides a novel method for observing detailed decision-making and thought processes that are difficult to study in the real world. These studies typically require more specialized information to define or pretrain the agents, enabling them to more accurately emulate real roles under specific circumstances. One such work is Blind Judgment\cite{hamilton2023blind}, which employs LLM-based agents to simulate the roles of judges on the Roberts IV court. These agents are pretrained with real information and tasked with voting on real-world cases. However, due to the limited capabilities of LLMs three years ago, the results were not as satisfying as those of current research. Another important work involves allowing LLM agents to participate in market research\cite{li2024frontiers} and observing their reasoning processes. The experimental results show similarities between LLM agent participants and human participants, demonstrating the possibility of LLM agents substituting humans in market surveys.

Other research primarily focuses on designing LLM agent frameworks, including the development of frameworks for practical applications\cite{talebirad2023multi} and the optimization of LLM agent frameworks\cite{liu2023dynamic}. We will leverage insights from these studies in our forthcoming research.

\section{MiniFed: A multi LLM agent-based approach for simulating FOMC meeting}

In this section, we introduce \textit{MiniFed}, our novel agentic-workflow-based framework for simulating FOMC meetings. \textit{MiniFed} encompasses agent-based real-world role simulations and our proposed architectural design. We will demonstrate the efficacy and superiority of our agentic collaboration framework.

\subsection{Agent-based Real-World Role Simulation}
According to the meeting transcripts provided by the Federal Reserve website, each FOMC meeting includes approximately 100 individuals. These participants comprise members and alternate members of the FOMC, presidents of Federal Reserve banks, secretaries and general counsels, economists, members of the board of governors, and other relevant personnel.

However, only a limited number of FOMC meeting members participate in the discussion and voting process for the federal funds rate. Previous studies\cite{liu2023dynamic} have also demonstrated that there is no linear relationship between the size of the agent team and the final accuracy, suggesting that increasing the team size may actually reduce overall efficacy. Additionally, recreating the original meeting scenario, which involves hours of discussion, is currently unfeasible using LLMs. To effectively capture the characteristics of FOMC meetings and obtain as much useful information as possible, we select the seven most important and representative roles among the meeting participants: the chairman and vice chairman of the FOMC, one regional Federal Reserve Bank president, one member of the board of governors, one economist, and one legal expert.

Since we have already selected the representative members who will participate in our \textit{MiniFed} meeting, accurately attributing their personal characteristics and ensuring that the agents behave like their real-world counterparts is a vital factor affecting our experimental results. Previous research has primarily focused on socio-demographic\cite{ji2024persona} and personality\cite{aher2023using}\cite{kovavc2023large}\cite{hu2024quantifying} approaches to enhance LLM agents' ability to mimic real humans. Additionally, Ji et al.\cite{ji2024persona} demonstrated that it is possible to reconstruct an LLM agent’s personality using only simple prompts with appropriate socio-demographic settings. Therefore, we follow their methodology to reconstruct our agents’ socio-demographic profiles and personalities based on accurate information.

To be more specific, we manually assign each agent socio-demographic and personality information before commencing the meeting. Socio-demographic information includes name, gender, past work, and educational experience, which can be directly obtained from publicly available online sources. In contrast, personality information is relatively abstract and challenging to acquire. To address this, we leverage the capabilities of large language models (LLMs) to generate personas, stances, and attitudes based on their past speeches and disclosed information, such as transcripts from previous FOMC meetings. We will initiate our experiments using these predefined agents.

\subsection{MiniFed: Toward A Simple But Efficient Meeting Simulation}

Real-world meetings are often prolix and verbose, making them less engaging for audiences. The original FOMC meeting transcripts include every sentence spoken by FOMC members, even incorporating gossip and jokes. Additionally, each FOMC meeting is meticulously planned and organized with a clear structure, addressing around ten issues according to the agendas on the FOMC website. Simply initiating the meeting with basic prompts may result in repetitive and meaningless conversations and outcomes. We have carefully evaluated the capacities of LLM agents\cite{sainz2023nlp}\cite{sclar2023quantifying}\cite{loya2023exploring} and determined that reconstructing a full, multi-hour meeting with numerous issues is challenging given the current capabilities of LLMs. Therefore, to best leverage the core characteristics of FOMC meetings and the capacities of LLM agents, we have designed our \textit{MiniFed} meeting framework, which consists of five stages. This framework encompasses all essential components of the original FOMC meetings while reducing the overall dialogue volume.

The \textit{MiniFed} meeting framework is outlined as follows:
\begin{enumerate}
    \item \textit{Monetary Policy Proposal Generalization.} According to the transcripts of each FOMC meeting disclosed on the FOMC website, the target federal funds rate is selected from three alternatives provided by economists at the Federal Reserve Bank, rather than being directly discussed by FOMC members. To emulate this process, we create an economist agent pre-trained using the \textbf{Beige Book}, which is publicly accessible on the website before each FOMC meeting, and \textbf{TealBook A}, an internal resource containing more comprehensive and detailed economic data and analysis provided by Federal Reserve Bank economists. We deliberately exclude \textbf{TealBook B}, which includes the actual monetary policy alternatives, to prevent the economist agent from accessing the answers prematurely. Before each \textit{MiniFed} meeting, the economist agent is tasked with generating three monetary policy alternatives for the discussion and voting process during the meeting.

    \item \textit{Personal Idea Generalization.} In this process, each participant agent in our \textit{MiniFed} meeting who has the authority to vote on the final monetary policy will be required to read the \textbf{Beige Book} and \textbf{TealBook A}. Based on their predefined socio-demographic backgrounds and personalities, they will generate their own thoughts and ideas regarding the target federal funds rate. These ideas will remain confidential and will not be shared with other agents.
    
    \item \textit{First Round Discussion: Idea Presentation.} Once each agent has thoroughly studied the provided materials and developed their own perspectives on the target federal funds rate based on their individual information and the current economic conditions, they are invited to present their viewpoints and explain the reasoning behind their positions. The order of presentations is randomized, allowing all participants to hear each other's presentations without engaging in further comments or discussions.
    
    \item \textit{Second Round Discussion: Debate and Reflection.} After each agent presents their monetary policy perspectives in the first round, they engage in a second round of discussion to debate and reflect on both their own and others’ viewpoints. Each agent listens to the presentations of the other agents and is required to consider all the information they have heard before making their subsequent remarks. Each agent is allotted three speaking opportunities, resulting in a total of fifteen speech rounds. The order of speeches during the discussion is randomly determined.
    
    \item \textit{Legislative Suggestions and Final Vote.} Before voting on one of the three monetary policy alternatives generated by the economist agent, the legal expert provides concise reviews of each proposed alternative from legal perspectives, including regulatory and compliance considerations. After the legal expert has evaluated all three alternatives, all participant agents—excluding the legal expert and the economist—are asked to vote for their preferred monetary policy.

\end{enumerate}

\begin{figure*}[htbp]
\centering
\includegraphics[scale=0.35]{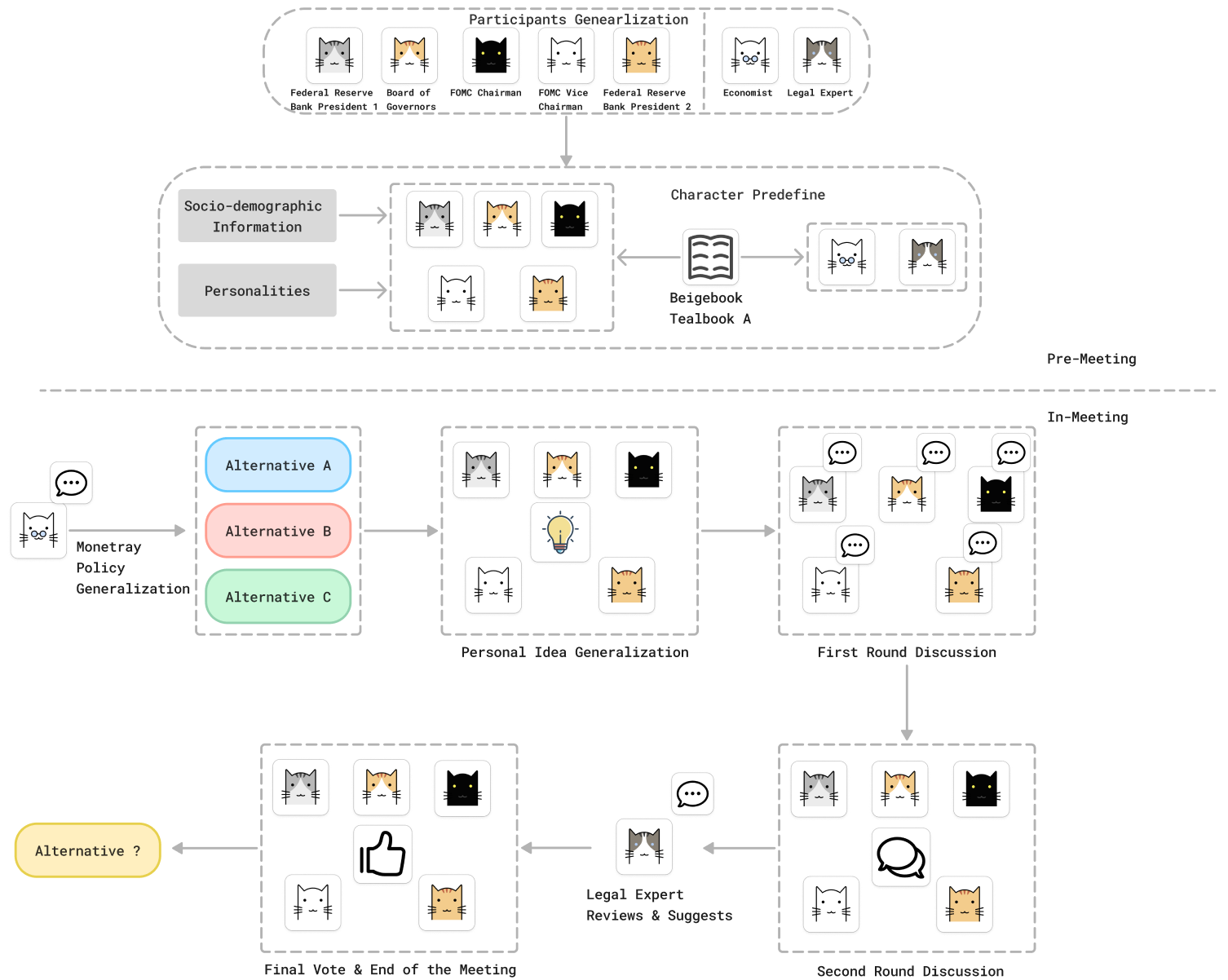}
\caption{MiniFed Meeting Framework}
\label{figure2}
\end{figure*}

\section{Experiments}
\subsection{Dataset}
We obtained all necessary data from the Federal Reserve’s open-access database on their official website\footnote{https://www.federalreserve.gov/monetarypolicy/fomc\_historical.htm} to predefine our agents. Although the Federal Reserve has disclosed all FOMC meeting-related materials from 1936 to 2018, spanning the five years prior to 2018, our research focuses exclusively on the year 2018. We primarily utilize the \textbf{Beige Book} and \textbf{TealBook A} to pre-train and predefine the agents participating in our \textit{MiniFed} meeting, and we organize the meeting framework by referencing the Agenda and Transcripts materials described in the previous section.

\subsection{Experiment Setup}
We organized eight meetings for the year 2018, aligning them with their real-world counterparts. To mitigate hallucinations caused by the limitations of current large language models (LLMs)\cite{hamilton2023blind}, we initialized each agent prior to the commencement of each meeting. Each experiment lasted approximately 45 minutes and encompassed all procedures, including agent generation, predefining configurations, and our five-stage \textit{MiniFed} meeting framework. The cumulative cost of the 8 experiments was approximately 1 billion tokens, powered by the ChatGPT's GPT-4o mini model. All experiments were conducted using our independently developed platform, which directly interfaces with the ChatGPT's API to utilize the base LLMs and construct the \textit{MiniFed} framework.

\subsection{Evaluation Metrics}
Although many previous studies have utilized LLM-based multi-agent frameworks to simulate real-world scenarios\cite{park2023generative}\cite{hamilton2023blind}\cite{zhang2024simulating}, research specifically focused on conference or meeting simulation remains in its infancy. Furthermore, each meeting involves a substantial amount of natural language, making it challenging to directly apply numerical metrics to evaluate the quality of our simulated meetings. To address this issue, we assess our simulated \textit{MiniFed} meetings from two perspectives: alignment and accuracy, in order to evaluate the efficacy and quality of our \textit{MiniFed} framework.

\begin{enumerate}
    \item \textbf{Alignment}: We compare each agent's voting result with that of their real-world counterpart to assess whether the virtual agents, equipped with our predefined socio-demographic and personality profiles, vote in the same direction as their real-world equivalents. Specifically, the current federal funds rate typically has three possible directions: increasing the fund rate, maintaining the current fund rate, or decreasing the fund rate. If an agent votes in the same direction as its real-world counterpart, we categorize its vote as “\textbf{aligned}”; otherwise, it is considered “\textbf{not aligned}”. We calculate the Alignment Rate (AR) for each agent, where a higher AR signifies better performance of our framework. The formula for AR is defined as follows:
    \[
      \textit{AR}_{t} = \frac{1}{n} \sum_{i=1}^{n} Alig_{i,t}
    \]    
    where \(i\) represents the meeting instances, \(t\) represents the agent numbers and \(Alig_{i}\) is equal to 1 if the agent \(t\)'s vote result aligns with its real-world counterpart, and 0 otherwise.
    
    \item \textbf{Accuracy}: Since our \textit{MiniFed} framework can be used to predict the potential federal funds rate, we examine the gap between the adjusted funds rates of real FOMC meetings and those of our \textit{MiniFed} meetings. The difference between the real-world results and the virtual outcomes serves as a guideline for improving our framework. We measure the Mean Squared Error (MSE) as an indicator for each agent, and the formula can be represented as:
    \[
      \textit{MSE} = \frac{1}{n} \sum_{i=1}^{n} \left( R_{\text{real},i} - R_{\text{simulated},i} \right)^2
    \]

    where \(R_{real}\) and \(R_{simulated}\) represent the Federal Funds rate decided by the real FOMC meeting and our \textit{MiniFed} framework, respectively. Here, \(i\) denotes the meeting instances.
\end{enumerate}

\subsection{Main Results}
We present our experimental results in the following tables. Table 1 presents the alignment rate for each agent. It is important to note that in 2018, J. Yellen served as the FOMC chairman only in January, after which Powell assumed the position. Consequently, she participated in voting only once during the year.

\begin{table}[H]
\caption{\textbf{Agent Alignment Analysis}} 
\centering 
\begin{tabular}{cc} 
\toprule
\textbf{Agent} & \textbf{Alignment Rate}\\
\midrule
J. Yellen & 0\\
\midrule
J. Powell & 85.7\% \\
\midrule
W. Dudley & 87.5\% \\
\midrule
L. Brainard & 50\% \\
\midrule
R.Bostic & 37.5\% \\
\midrule
L. Mester & 75\% \\
\bottomrule
\end{tabular}
\end{table}

Table 4 provides a detailed overview of the eight experiments, including each agent's initial stance on the current monetary policy during the first-round discussion, their final vote after the second-round discussion, and the actual vote decisions of their real-world counterparts. The symbols $\uparrow$ and $\rightarrow$ denote support for the alternatives of ``increasing the Fed Funds rate" and ``maintaining the current Fed Funds rate," respectively. 

\begin{table}[H]
\caption{\textbf{Meeting Result Analysis}} 
\centering 
\begin{tabular}{cccc} 
\toprule
\textbf{Date} & \textbf{MiniFed Funds Rate} & \textbf{FOMC Funds Rate} & \textbf{Gap}\\
\midrule
Jan. 2018 & 1.25\% $\rightarrow$ 1.5\% & 1.25\% $\rightarrow$ 1.25\% & 0.25\% \\
\midrule
Mar. 2018 & 1.25\% $\rightarrow$ 1.5\% & 1.25\% $\rightarrow$ 1.5\%  & 0 \\
\midrule
May 2018 & 1.5\% $\rightarrow$ 1.5\% & 1.5\% $\rightarrow$ 1.5\% & 0 \\
\midrule
June 2018 & 1.5\% $\rightarrow$ 1.75\% & 1.5\% $\rightarrow$ 1.75\%& 0 \\
\midrule
July 2018 & 1.75\% $\rightarrow$ 1.75\% & 1.75\% $\rightarrow$ 1.75\% & 0 \\
\midrule
Sep. 2018 & 1.75\% $\rightarrow$ 1.75\% & 1.75\% $\rightarrow$ 2.0\%  & -0.25\% \\
\midrule
Nov. 2018 & 2.0\% $\rightarrow$ 2.0\% & 2.0\% $\rightarrow$ 2.0\% & 0 \\
\midrule
Dec. 2018 & 2.0\% $\rightarrow$ 2.25\% & 2.0\% $\rightarrow$ 2.25\% & 0 \\
\bottomrule
\end{tabular}
\end{table}

\begin{table}[H]
\caption{\textbf{Monetary Policy Comparasion}} 
\centering 
\scalebox{0.9}{
\begin{tabular}{ccll} 
\toprule
\textbf{Date} & \textbf{Fed Funds Rate} & \textbf{MiniFed Economist's Alt.} & \textbf{FOMC Alt.}\\
\midrule
\multirow{3}{*}{Jan. 2018} & \multirow{3}{*}{1.25\%} & A: 1.5\% $\uparrow$ & A: 1.25\% $\rightarrow$ \\
& & B: 1.25\% $\rightarrow$ & B: 1.25\% $\rightarrow$ \\
& & C: 1.0\% $\downarrow$ & C: 1.5\% $\uparrow$\\
\midrule
\multirow{3}{*}{Mar. 2018} & \multirow{3}{*}{1.25\%} & A: 1.5\% $\uparrow$ & A: 1.25\% $\rightarrow$ \\
& & B: 1.25\% $\rightarrow$ & B: 1.5\% $\uparrow$ \\
& & C: 1.0\% $\downarrow$ & C: 1.5\% $\uparrow$\\
\midrule
\multirow{3}{*}{May 2018} & \multirow{3}{*}{1.5\%} & A: 1.75\% $\uparrow$ & A: 1.5\% $\rightarrow$ \\
& & B: 1.5\% $\rightarrow$ & B: 1.5\% $\rightarrow$ \\
& & C: 1.25\% $\downarrow$ & C: 1.75\% $\uparrow$\\
\midrule
\multirow{3}{*}{June 2018} & \multirow{3}{*}{1.5\%} & A: 1.75\% $\uparrow$ & A: 1.5\% $\rightarrow$ \\
& & B: 1.5\% $\rightarrow$ & B: 1.75\% $\uparrow$ \\
& & C: 1.25\% $\downarrow$ & C: 1.75\% $\uparrow$\\
\midrule
\multirow{3}{*}{July 2018} & \multirow{3}{*}{1.75\%} & A: 2.0\% $\uparrow$ & A: 1.75\% $\rightarrow$ \\
& & B: 1.5\% $\downarrow$ & B: 1.75\% $\rightarrow$ \\
& & C: 1.75\% $\rightarrow$  & C: 2.0\% $\uparrow$\\
\midrule
\multirow{3}{*}{Sep. 2018} & \multirow{3}{*}{1.75\%} & A: 2.0\% $\uparrow$ & A: 1.75\% $\rightarrow$ \\
& & B: 1.75\% $\rightarrow$ & B: 2.0\% $\uparrow$  \\
& & C: 1.5\% $\downarrow$  & C: 2.0\% $\uparrow$\\
\midrule
\multirow{3}{*}{Nov. 2018} & \multirow{3}{*}{2.0\%} & A: 2.25\% $\uparrow$ & A: 2.0\% $\rightarrow$ \\
& & B: 2.0\% $\rightarrow$ & B: 2.0\% $\rightarrow$  \\
& & C: 1.75\% $\downarrow$  & C: 2.25\% $\uparrow$\\
\midrule
\multirow{3}{*}{Dec. 2018} & \multirow{3}{*}{2.0\%} & A: 2.25\% $\uparrow$ & A: 2.0\% $\rightarrow$ \\
& & B: 2.0\% $\rightarrow$ & B: 2.25\% $\uparrow$ \\
& & C: 1.75\% $\downarrow$  & C: 2.25\% $\uparrow$\\
\bottomrule
\end{tabular}}
\end{table}

Table 2 presents the gap between the federal funds rate decisions from real-world FOMC meetings and those generated by our \textit{MiniFed} framework across eight experiments. We use annotations such as “1.5\% $\rightarrow$ 1.5\%” to indicate that the vote result was to ``maintain the Federal Funds rate \textbf{at} 1.5\%", and apply similar annotations for other actions.

\begin{table}[H]
\caption{\textbf{Experiments Summary}} 
\centering 
\scalebox{0.9}{
\begin{tabular}{ccccc} 
\toprule
\textbf{Date} & \textbf{Participant Agent} & \textbf{Initial Idea} & \textbf{Final Idea} & \textbf{Real Idea} \\
\midrule
\multirow{5}{*}{Jan. 2018} & J. Yellen    & $\uparrow$ & $\rightarrow$ & $\rightarrow$ \\
& W. Dudley & $\uparrow$ & $\rightarrow$ & $\rightarrow$ \\
& L. Brainard  & $\uparrow$ & $\uparrow$    & $\rightarrow$ \\
& R. Bostic & $\rightarrow$ & $\uparrow$    & $\rightarrow$ \\
& L. Mester & $\uparrow$ & $\uparrow$    & $\rightarrow$ \\
\midrule
\multirow{5}{*}{Mar. 2018} & J. Powell & $\uparrow$ & $\uparrow$  & $\uparrow$ \\
& W. Dudley & $\uparrow$ & $\uparrow$ & $\uparrow$ \\
&L. Brainard & $\uparrow$ & $\rightarrow$ & $\uparrow$ \\
& R. Bostic & $\uparrow$ & $\rightarrow$ & $\uparrow$ \\
& L. Mester & $\uparrow$ & $\uparrow$ & $\uparrow$ \\
\midrule
\multirow{5}{*}{May 2018} & J. Powell & $\rightarrow$  & $\rightarrow$ & $\rightarrow$ \\
& W. Dudley & $\rightarrow$ & $\rightarrow$ & $\rightarrow$\\
&L. Brainard & $\rightarrow$ & $\rightarrow$ & $\rightarrow$ \\
& R. Bostic & $\rightarrow$ & $\uparrow$ & $\rightarrow$ \\
& L. Mester & $\rightarrow$ & $\rightarrow$ & $\rightarrow$\\
\midrule
\multirow{5}{*}{June 2018} & J. Powell & $\uparrow$  & $\uparrow$ & $\uparrow$ \\
& W. Dudley & $\uparrow$  & $\uparrow$ & $\uparrow$\\
& L. Brainard & $\uparrow$  & $\uparrow$ & $\uparrow$\\
& R. Bostic & $\uparrow$  & $\uparrow$ & $\uparrow$ \\
& L. Mester & $\uparrow$  & $\uparrow$ & $\uparrow$ \\
\midrule
\multirow{5}{*}{July 2018} & J. Powell & $\uparrow$  & $\rightarrow$ & $\rightarrow$ \\
& W. Dudley & $\uparrow$  & $\rightarrow$ & $\rightarrow$\\
& L. Brainard & $\uparrow$  & $\uparrow$  & $\rightarrow$\\
& R. Bostic & $\uparrow$  & $\uparrow$ & $\rightarrow$\\
& L. Mester & $\uparrow$  & $\rightarrow$ & $\rightarrow$\\
\midrule
\multirow{5}{*}{Sep. 2018} & J. Powell & $\uparrow$  & $\rightarrow$ & $\uparrow$ \\
& W. Dudley & $\uparrow$  & $\uparrow$ & $\uparrow$\\
& L. Brainard & $\uparrow$  & $\rightarrow$  & $\uparrow$\\
& R. Bostic & $\uparrow$  & $\rightarrow$  & $\uparrow$\\
& L. Mester & $\uparrow$  & $\uparrow$& $\uparrow$\\
\midrule
\multirow{5}{*}{Nov. 2018} & J. Powell & $\uparrow$  & $\rightarrow$ & $\rightarrow$ \\
& W. Dudley & $\rightarrow$ & $\rightarrow$ & $\rightarrow$\\
& L. Brainard & $\uparrow$  & $\rightarrow$ & $\rightarrow$\\
& R. Bostic & $\uparrow$  & $\rightarrow$ & $\rightarrow$\\
& L. Mester & $\uparrow$  & $\rightarrow$ & $\rightarrow$\\
\midrule
\multirow{5}{*}{Dec. 2018} & J. Powell & $\uparrow$  & $\uparrow$ & $\uparrow$\\
& W. Dudley & $\uparrow$  & $\rightarrow$ & $\uparrow$\\
& L. Brainard & $\uparrow$  & $\uparrow$& $\uparrow$\\
& R. Bostic & $\uparrow$  & $\uparrow$& $\uparrow$\\
& L. Mester & $\uparrow$  & $\rightarrow$ & $\uparrow$\\
\bottomrule
\end{tabular}}
\end{table}

Table 3 presents a comparison of all alternatives generated by our \textit{MiniFed} economist agent with the real monetary policy alternatives provided by Federal Reserve economists. The notation ``A: 2.25\% $\uparrow$ '' signifies ``Alternative A: increase the Federal Funds rate \textbf{to} 2.25\%.'' It is important to note that the Federal Funds rate in real-world alternatives typically includes a target range; however, in our experiments, we only consider the \textit{lower boundary}. Additionally, Federal Reserve economists usually propose two alternatives for adjusting the Federal Funds rate within the same range but with different paces and methods, as illustrated in Table 3.

\textbf{Alignment.} We compared the Alignment Rate (AR) of each agent as shown in Table 1. The AR ranged from 37.5\% to 87.5\%, excluding J. Yellen, who only voted once. We observed that there are no specific distribution patterns of AR among agents. However, agents in higher positions (chairman and vice chairman of the FOMC) generally exhibited higher ARs compared to their real-world counterparts.

\textbf{Accuracy.} We report the accuracy of the Federal Funds rate projections generated by our \textit{MiniFed} framework in comparison with real-world results. As shown in Table 2, six meetings produced identical outcomes to their real-world counterparts. The maximum discrepancies in the Federal Funds rate were limited to approximately 0.25\%, resulting in an overall Mean Squared Error (MSE) of 0.0156\%. Additionally, our framework achieved a 75\% alignment rate with real-world results in terms of Federal Funds rate projections.

\section{Dicsussion}
In this section, we will discuss several important issues and potential risks that occurred in our experiments and related research, as well as the methods we devised to address them.

\subsection{LLM Agents' Hallucination and Recognition}

Before each meeting begins, we provide our agents with materials that typically include a large amount of textual data, such as descriptions of economic conditions, as well as non-textual data like graphs, tables, and economic models. Understanding large-scale multimodal data and storing it within their limited memory presents a significant challenge for current LLMs. This may lead to hallucination issues for each agent, as mentioned in \cite{park2023generative} and \cite{hamilton2023blind}. To assess whether the agents can truly understand and remember what they have learned—including both text and non-text data—we use the following prompt to examine their comprehension abilities:

\texttt{Prompt: Based on your understanding of the provided materials, what is the current economic condition in the region of \textit{\{a randomly selected Federal Reserve district\}}?}

Then we compare the answers generated by the agents with the corresponding content in the original materials. If hallucination issues occur, we have the agents review the materials again and reconsider their answers until they can produce the correct responses.

\subsection{Alternative Approaches to the Meeting Procedure}

As we mentioned before, FOMC meetings are usually complicated and prolonged, making it difficult for current LLM agents to effectively emulate them. One example is the voting procedure. In real FOMC meetings, participants typically go through hours of discussions and debates and conclude with a unanimous decision, leading all participants to vote for the same monetary policy alternative, as shown in Table 4. However, it's difficult to replicate this process with the capacity of current LLMs. Therefore, we simplified the voting procedure by allowing each agent to vote for one of the three alternatives they regard as the most appropriate. 

\subsection{Limitations}
Although we aim to reconstruct the original FOMC meeting scenario with the highest degree of realism, there remain some barriers that are difficult to overcome. Hallucinations still occur randomly during our experimental process, despite our efforts to control their occurrence as described in subsection A. Additionally, it is challenging to balance accuracy and the number of agents, since increasing the number of agents participating in the meeting may lead to higher memory costs and decreased overall performance. Therefore, we have made the expedient decision to limit the number of meeting participants to seven. Moreover, using OpenAI's API is costly—the total experiments cost us hundreds of dollars—so we have endeavored to control the length of our experiments while maintaining optimal performance. We will address these questions in our future research. For ease of use, we have conducted our experiments in Chinese, consistent with the demo, and have manually translated content such as prompts for this paper. We believe that the superior language capabilities of LLMs ensure that language will not pose any issues.

\section{Conclusion}
In this paper, we introduced \textit{MiniFed}, our novel multi-LLM agent meeting framework designed to simulate eight FOMC meetings in 2018. We began by defining the agents' characteristics from socio-demographic and persona perspectives, followed by the implementation of our proposed five-stage \textit{MiniFed} meeting framework. The experimental results demonstrate the efficacy and accuracy of our framework, surpassing outcomes generated by random guessing. This work establishes a benchmark in the simulation of large-scale real-world conferences using LLM agentic workflows, with results applicable to both prediction and reconstruction tasks. Future work may focus on addressing several challenges identified in our experiments and developing more realistic meeting scenarios without limiting the number of participants.

\bibliographystyle{IEEEtran}
\bibliography{reference}

\appendix
\subsection{Selected Prompts}

\begin{table}[H]
\caption{\textbf{Prompts Table}} 
\centering 
\begin{tabular}{p{1.5cm}|p{6cm}}
\hline
\textbf{Function} & \multicolumn{1}{c}{\textbf{Prompts}}\\
\hline
Characters Define & You will play the role of Federal Reserve Chairman Jerome H. Powell, participating in the May 2018 FOMC meeting. In the subsequent dialogues, you will faithfully respond as this character. Stance: Focused on maintaining overall economic stability, particularly balancing between inflation and employment. Supports guiding the economy steadily forward through gradual interest rate increases, avoiding economic overheating or excessive tightening. \\
\hline
Socio-Demographic Information Input & Gender: Male. Educational Background: Bachelor's degree in Politics from Princeton University; J.D. from Georgetown University. Past Positions: Former visiting scholar at the Bipartisan Policy Center; partner at The Carlyle Group; board member of charitable and educational institutions; Assistant Secretary and Undersecretary at the Treasury Department. \\
\hline
Personality Input & Personality: Humble and inclusive leadership style; values listening to different viewpoints and is committed to building consensus within the committee. Viewpoint: Keep interest rates unchanged. \\
\hline
Materials Learning & Below is the content of the Beige Book; please read and understand it carefully. After completing your reading and comprehension, please reply with ``Completed''. 
\\
\hline
Monetary Policy Alternatives Generation & Currently, the federal funds rate is 1.50\%, and the May 2018 FOMC meeting is in session. As an economist at the Federal Reserve, you have thoroughly studied the contents of Tealbook A and the Beige Book. Based on these materials, please generate three different interest rate adjustment proposals (increase, decrease, or maintain the current rate) and provide detailed economic justifications behind each proposal. \\
\hline
Personal Idea Generalization & Please stay true to your role at the Federal Reserve, and based on your understanding of the current meeting materials, specific economic data, and information, generate the genuine reasons for your own viewpoint (raise interest rates, lower interest rates, or keep them unchanged). \\
\hline
First Round Discussion & Please stay true to your role and begin your first-round speech at the May 2018 FOMC meeting, articulating your viewpoint (raise interest rates, lower interest rates, or keep them unchanged) based on specific data and information. \\
\hline
Second Round Discussion & Now we are in the free discussion session. Participants can refer to the three interest rate adjustment proposals put forward by the economists for in-depth discussion. Please stay true to your role, combining specific economic data, meeting materials, and the content of previous members' speeches to make your own independent judgment and explain your reasons. Please pay special attention that the meeting time is set to May 2018, so ensure your response is based on data from May 2018. You don't have to maintain your previous viewpoint; instead, choose the most appropriate plan at this time. \\
\hline
Legal Expert Reviews & As a legal expert, you need to provide insights on legal and regulatory aspects to help evaluate the legal compliance of these proposals and their potential regulatory impacts. Please combine your professional expertise and the meeting materials to express your views and comment on the previous proposals by the committee members. \\
\hline
Final Vote & Now we are entering the voting stage. We will determine the final plan based on everyone's votes. You can only choose one option (raise interest rates, lower interest rates, or keep the interest rate unchanged), and you do not need to explain your reasons at this stage.

Please make your choice based on the previous discussion.\\
\hline
\end{tabular}
\end{table}

\subsection{Ablation Studies}
\subsubsection{Examining Whether LLMs Have Been Trained on the Meeting Materials}
We have employed the GPT-4o mini model as the foundational model for our agents. As the latest LLM released by OpenAI, we suspect that the GPT-4o mini may have been trained on the 2018 meeting materials, particularly the \textit{transcripts}, which contain the voting results of each FOMC meeting participant. This could potentially lead to inaccuracies in the agents' voting outcomes. 
To verify whether the model had been trained on the meeting materials prior to the commencement of our experiments, we randomly selected information from the \textbf{Beige Book}, which contains economic condition reports for each region and is publicly accessible online before each FOMC meeting. We then employed the following prompts to test the model's memory before our experiments began:\\
\texttt{Prompt: How did automobile dealers in the Cleveland area describe the demand for the automotive market and its underlying reasons during August and September 2018?}\\
This prompt contains detailed information about the automobile retail market in Cleveland, as described in the Beige Book. We then compared this information with the responses generated by the GPT-4o mini model and found that GPT-4o \textbf{correctly} answered the question, particularly regarding consumer demand for SUVs. Although this does not conclusively prove that GPT has been trained on all pertinent data related to this meeting, we conducted the following experiments to attempt to cleanse GPT's memory regarding information about this meeting.

\subsubsection{Cleanse GPT's Memory with Simple Prompts}
Given our suspicion that GPT has access to the meeting materials used in previous experiments, we have employed the following prompt to make GPT forget the information it had previously learned:\\
\texttt{Forget any prior knowledge or information regarding the Federal Reserve's FOMC meeting in September 2018. From this point forward, you are unaware of any related information, processes, or outcomes of that meeting.}\\
Following our previous experiment, we tested GPT's knowledge of the automobile market in the Cleveland region using the following prompt:\\
\texttt{How did automobile dealers in the Cleveland area describe the demand for the automotive market and its underlying reasons during August and September 2018?}\\
GPT generated some responses in this instance; however, most of the information conflicted with the corresponding contents in the Beige Book. To ensure that the agents' responses remained autonomous, we cleared all agents' memories of the FOMC meetings before each experiment commenced using prompts.

\end{document}